\documentclass[superscriptaddress,floatfix,amsmath,amssymb,aps,prx,reprint]{revtex4-2}
\usepackage{physics}
\usepackage{bbold}
\usepackage{natbib}    
\usepackage{color}         
\usepackage{graphicx}      
\usepackage{epsfig}          
\usepackage{bm}            
\usepackage{amsmath}
\usepackage{thumbpdf}
\usepackage{hyperref} 
\usepackage{placeins}
\usepackage{xcolor}
\usepackage{multirow}
\usepackage{caption}
\usepackage[normalem]{ulem}

\usepackage[utf8]{inputenc}

\begin{document}

\title{Large spin splitting at ferromagnetic surfaces of bulk antiferromagnets}
\author{William A. Schaarman}
\affiliation{Department of Physics and Astronomy, Chalmers University of Technology, SE-412 96 Gothenburg, Sweden}
\author{Sophie F. Weber}
\affiliation{Department of Physics and Astronomy, Chalmers University of Technology, SE-412 96 Gothenburg, Sweden} 
\date{\today}
\begin{abstract}
We use density functional theory and model Hamiltonians to reveal large spin splitting of bands localized at ferromagnetic surfaces of bulk antiferromagnets (AFMs). There is great interest in material platforms combining the robustness and ultrafast dynamics of AFMs with large, functional spin splitting which is often restricted to ferromagnets (FMs). Here, we show that a subset of AFM \textit{surfaces} which have symmetry-allowed magnetization can host large spin splitting via bulk degeneracy lifting of sublattice-resolved exchange splittings. We find that the spin splitting is maximized for two ferromagnetic surface motifs: terminations with single uncompensated magnetic sublattices, and two-sublattice surfaces whose sublattices are magnetically compensated in the bulk, but acquire distinct crystal field environments via surface truncation. The latter case can yield FM-like spin splitting magnitudes while also having small uncompensated magnetization. We confirm these predictions with first-principles calculations of $\mathrm{Cr_2O_3}$ and $\mathrm{FeF_2}$, finding splittings as large as $1~\mathrm{eV}$ depending on the surface in question. Our findings point to intrinsic surface symmetry breaking as a route to large, functional spin splitting in an expanded range of AFM materials.
\end{abstract}

\maketitle




 Antiferromagnets (AFMs) have potential for use in next-generation logic and storage devices. Their staggered magnetization, or N\'{e}el vector, can encode information in a bit-like manner, their vanishing net magnetization gives robustness to stray magnetic fields, and their $\sim\mathrm{THz}$ magnetization dynamics make them promising for ultrafast processing applications~\cite{chenEmergingAntiferromagnetsSpintronics2024}. However, the same symmetries that guarantee a vanishing net magnetization in many AFMs (the product of inversion and time-reversal symmetry, $\mathcal{I}\mathcal{T}$, or time-reversal coupled to a fractional lattice translation, $\mathcal{T}t$) prohibit the spin split electronic structure typical to ferromagnets (FMs). Spin split bands enable generation of a spin-polarized current, which is fundamental to numerous spintronic device architectures~\cite{khvalkovskiyBasicsPrinciplesSTTMRAM, ralphSpinTransferTorques2008, songAltermagnetsNewClass2025}. Hence, materials combining AFM robustness with spin split bands are highly desirable.

Several routes to spin splitting in AFMs have been proposed. Examples include altermagnets (AMs)~\cite{smejkalEmergingResearchLandscape2022}, AFMs which host momentum-dependent spin splitting due to intrinsically broken bulk $\mathcal{I}\mathcal{T}$ and $\mathcal{T}t$ symmetries; $\mathcal{I}\mathcal{T}$ symmetry breaking via applied electric fields~\cite{taoLayerHallDetection2024}, and symmetry breaking through atomic substitution in Janus structure AFMs~\cite{guoSymmetrybreakingInducedTransition2025, zhuMultipiezoEffectAltermagnetic2023}. However, a persistent challenge with these and other platforms is that the associated spin splittings are generally small compared to the $1$-$2~\mathrm{eV}$ splittings of conventional FMs~\cite{eastmanExperimentalExchangeSplitEnergyBand1980}. Electric-field- and Janus-based mechanisms produce maximum splittings on the order of hundreds of $\mathrm{meV}$~\cite{taoLayerHallDetection2024, luSpinsplittingRoomtemperatureJanus2025}. Excitingly, $\mathrm{eV}$-scale spin splittings have been reported in a few AMs, notably a $0.96~\mathrm{eV}$ spin splitting in $\mathrm{CrSb}$~\cite{yangThreedimensionalMappingAltermagnetic2025}, but these are exceptional cases with most proposed or experimentally discussed AMs exhibiting smaller splittings. In this work, similarly to Ref.~\cite{yangThreedimensionalMappingAltermagnetic2025}, we define spin splitting as the maximum energy difference between adjacent opposite-spin bands in the energy window of interest.\\
\indent In this Letter, we propose an alternative paradigm to achieve large spin splitting: \textit{surfaces} of bulk AFMs with a net surface magnetization. All surfaces inherently break $\mathcal{I}\mathcal{T}$ symmetry, and if specific additional bulk symmetries are broken, a surface-localized ferromagnetism (``surface magnetization") can emerge~\cite{belashchenkoEquilibriumMagnetizationBoundary2010,weberSurfaceMagnetizationAntiferromagnets2024}. While this surface magnetization, unlike bulk ferromagnetism, has negligible detrimental effect the robustness of bulk AFMs, especially for thick samples, it generically breaks $\mathcal{T}t$ symmetry and should be accompanied by an exchange interaction, leading to spin splitting of the surface-projected band structure. The real-space properties of such ferromagnetic surfaces have been investigated in numerous theoretical and experimental works\cite{appelNanomagnetismMagnetoelectricGranular2019,weberSurfaceMagnetizationAntiferromagnets2024,kosubAllElectricAccessMagneticFieldInvariant2015,pylypovskyiSurfaceSymmetryDrivenDzyaloshinskiiMoriyaInteraction2024}, but the nature of their corresponding reciprocal space electronic structure has yet to be explored.\\ 
\indent Our main finding is that a subset of the same mechanisms of symmetry-breaking that enable surface magnetization generically lead to large surface spin splittings. Using model Hamiltonians, we show that the magnitude of spin splitting depends on an interplay between crystal field (CF) energies and exchange splitting (EX) of individual surface atoms. Two ferromagnetic surface motifs are especially favorable: terminations with a single uncompensated magnetic sublattice, and two-sublattice surfaces where atomic truncation makes the sublattice crystal field environments inequivalent. In the latter case, CF-mediated shifts of oppositely polarized sublattice bands can enable $\mathrm{eV}$-scale splitting despite the small magnitude of uncompensated surface magnetization. In contrast, when the surface magnetization arises from relativistic canting of symmetry-connected sublattices, the associated spin splitting is small.\\
\begin{figure*}
    \centering
    \includegraphics[width=6.8in]{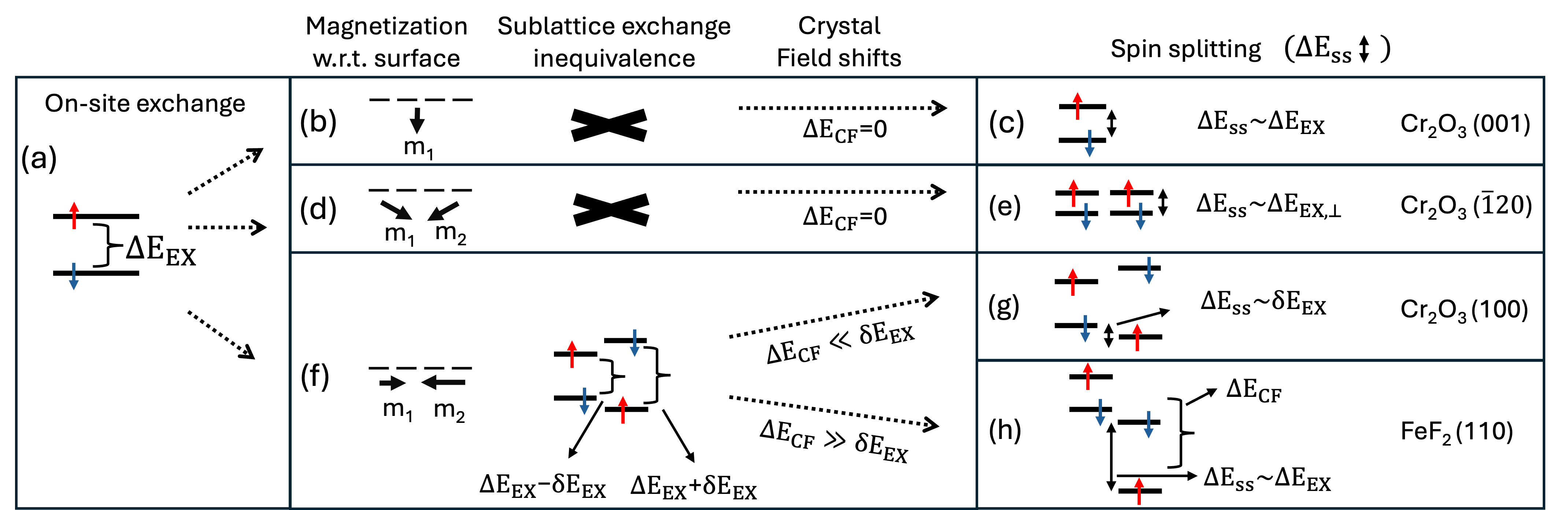}
    \caption{A schematic overview of distinct contributions to spin-splitting along with the relevant representative AFM facets studied in this work. $\mathbf{m_1}$ and $\mathbf{m_2}$ refer to distinct magnetic sublattices.}
    \label{fig:splitting}
\end{figure*}
We demonstrate these mechanisms with density functional theory (DFT) calculations of surface-projected band structures for representative facets of $\mathrm{Cr_2O_3}$ and $\mathrm{FeF_2}$. $\mathrm{Cr_2O_3}$ is a magnetoelectric AFM with $\mathcal{I}\mathcal{T}$ symmetry that enforces spin degeneracy in the bulk band structure, while $\mathrm{FeF_2}$ is a bulk altermagnet. Across the studied surfaces, we find spin splittings ranging from $\sim10~\mathrm{meV}$ to $\sim~1\mathrm{eV}$, with the magnitude controlled by surface symmetry and termination, consistent with the model. Our results establish surface symmetry breaking as a general route to large spin-split bands in an extended range of bulk AFMs. 

\textit{Theory} -- First, we review how two-dimensional magnetization can arise at surfaces of bulk AFMs. Given the bulk magnetic space group, the magnetic space group of a flat surface with a specific orientation defined by the normal vector $\mathbf{\hat{n}}$ is the subgroup consisting of all translation-symmetry pairs $(\mathbf{r},R)$ of the bulk group that have the following two properties: $R\mathbf{\hat{n}}=\mathbf{\hat{n}}$ and $\mathbf{r} \perp \mathbf{\hat{n}}$. If ferromagnetism is permitted in the corresponding surface point group, a net surface magnetization can emerge even when magnetization is prohibited in the bulk~\cite{weberSurfaceMagnetizationAntiferromagnets2024}. 

This magnetization affects the surface band structure by permitting an exchange interaction which splits the energies of the spin-up and spin-down states~\cite{taoInsulatortoconductorTransitionDriven2020, takenakaMagnetoelectricControlTopological2019}. It is useful to consider the exchange splitting of each distinguishable magnetic sublattice, where a sublattice is defined as the set of sites having the same magnitude and direction of magnetization via being connected through a surface group symmetry operation. To simplify the model, we consider on-site exchange splitting of a single representative $d$ orbital (Fig.~\ref{fig:splitting}(a)). The resulting contribution to the Hamiltonian is 
\begin{equation} \label{Hamiltonian,on-site}
    H=\frac{\Delta E_{EX}}{2}(\hat{\mathbf{m}}\cdot\sigma),
\end{equation}
where $\sigma=(\sigma_x,\sigma_y,\sigma_z)$ is the vector of Pauli matrices denoting spin degrees of freedom, $\Delta E_{EX}$ is the on-site exchange splitting and $\hat{\mathbf{m}}$ is a unit vector parallel to the sublattice magnetization direction.

Depending on how surface magnetization arises, additional energy terms can contribute to the overall spin-split band structure. First, we consider the  most intuitive case: a surface with a single uncompensated magnetic sublattice (Fig.~\ref{fig:splitting}(b)). The spin splitting $\Delta E_{ss}$ for this surface derives exclusively from the on-site splitting of the single sublattice via Eq.~\ref{Hamiltonian,on-site}, with $\hat{\mathbf{m}}$ along the direction of the Néel vector (Fig.~\ref{fig:splitting}(c)). In this and all other figures we label positive magnitudes of the energy difference. Note that while the spin splitting for the single band toy model in Eq.~\ref{Hamiltonian,on-site} and Fig.~\ref{fig:splitting}(c) is precisely $\Delta E_{EX}$, superposition of multiple non-degenerate exchange-split bands in a realistic band structure can lead to $\Delta E_{ss}$ which is reduced but on the same scale as $\Delta E_{EX}$ (we will see this is the case for the uncompensated $(001)$ surface of $\mathrm{Cr_2O_3}$).

Next, we consider the more subtle case of so-called ``induced" surface magnetization, discussed extensively in Refs.~\cite{weberSurfaceMagnetizationAntiferromagnets2024, pylypovskyiSurfaceSymmetryDrivenDzyaloshinskiiMoriyaInteraction2024}, which emerges on atomic planes that are magnetically compensated in the bulk. Such surfaces are composed of pairs of oppositely pointed magnetic sublattices; we consider a single sublattice pair with one exchange-split $d$ orbital per sublattice for simplicity.

There are two ways that surface magnetization can arise in such a model. If a symmetry operation in the surface group connects the two opposite-pointing sublattices, their magnetization magnitudes must remain the same. Thus, surface magnetization can only arise via a spin canting perpendicular to the Néel vector, with equal canting magnitudes on each sublattice (Fig.~\ref{fig:splitting}(d)). The Hamiltonian in this case is 
 \begin{equation} \label{Hamiltonian-120}
    H=\frac{\Delta E_{EX,\perp}}{2} \sigma_x\tau_0
\end{equation}
where $\Delta E_{EX,\perp}$ is the projection of $\Delta E_{EX}$ along the canting direction, which we take as $x$, and $\tau$ are Pauli matrices describing the sublattice degree of freedom (and $\tau_0$ is the identity matrix). The associated spin splitting is $\Delta E_{ss}\sim\Delta E_{EX,\perp}$ (Fig.~\ref{fig:splitting}(e)). As we shall see in our DFT calculations, $\Delta E_{ss}$ associated with Eq.~\ref{Hamiltonian-120} is generally small due to the small projections of the sublattice magnetic moments along the surface normal.

Induced surface magnetization can also arise when the surface group breaks symmetries which connect opposite sublattices in the bulk. In this case, their magnetization magnitudes become inequivalent, generically leading to a net surface magnetization parallel to the bulk Néel vector (Fig.~\ref{fig:splitting}(f), left). Depending on the specific surface symmetry, an additional perpendicular component from sublattice canting may occur, but since the effect will be similar to Eq.~\ref{Hamiltonian-120}, we neglect it here. 

The breaking of all sublattice-connecting symmetries has two consequences for the surface spin splitting. First, on-site exchanges of the two sublattices can now acquire different magnitudes ((Fig.~\ref{fig:splitting}(f), center) , which we write as $\Delta E_{EX}+\delta E_{EX}$ and $-\Delta E_{EX}+\delta E_{EX}$ ($|\Delta E_{EX}|>|\delta E_{EX}|$ ensures opposite signs). Secondly, once the sublattices are not connected by symmetry, the crystal fields set by their ligand environments no longer have to be equivalent. This can result in a rigid shift between energies of a given pair of $d$ bands belonging to opposite surface sublattices. We denote this additional shift as $\Delta E_{CF}$. Assuming the moments lie fully along the Néel vector ($z$-axis), the resulting Hamiltonian is given by

\begin{equation} \label{SpecificHamiltonian}
        H =\frac{\Delta E_{EX}}{2}\sigma_z \tau_z
        + \frac{\delta E_{EX}}{2}  \sigma_z \tau_0 +
        \frac{\Delta E_{CF}}{2} \sigma_0 \tau_z.
\end{equation}
Figs.~\ref{fig:splitting}(g)-(h) illustrate resulting spin-split band structures for limiting cases of the model in Eq.~\ref{SpecificHamiltonian}: $\delta E_{EX}>>\Delta E_{CF}$ and $\Delta E_{CF}>>\delta E_{EX}$ respectively. When $\delta E_{EX}>>\Delta E_{CF}$, $\Delta E_{ss}$ either corresponds to the small sublattice exchange \textit{difference} $\delta E_{EX}$, or to the smaller on-site exchange $\Delta E_{EX}-\delta E_{EX}$, depending on the energy window of interest. We label $\Delta E_{ss}$ in Fig.~\ref{fig:splitting}(g) corresponding to $\delta E_{EX}$, as this is the relevant splitting in the window we consider for $\mathrm{Cr_2O_3}$. In Fig.~\ref{fig:splitting}(h), on the other hand, the large $\Delta E_{CF}$ isolates the sublattice-projected exchange splittings, so adjacent opposite-spin bands corresponding to $\Delta E_{ss}$ are set by the larger-scale $\Delta E_{EX}$ throughout the spectrum. Thus, surfaces with a large $\Delta E_{CF}$ are the most robust two-sublattice motif for large spin splitting. We now discuss our ab-initio results to demonstrate realistic surfaces which can lead to the limiting cases in Fig.~\ref{fig:splitting}.

\begin{figure}
    \centering    \includegraphics[width=3.4in]{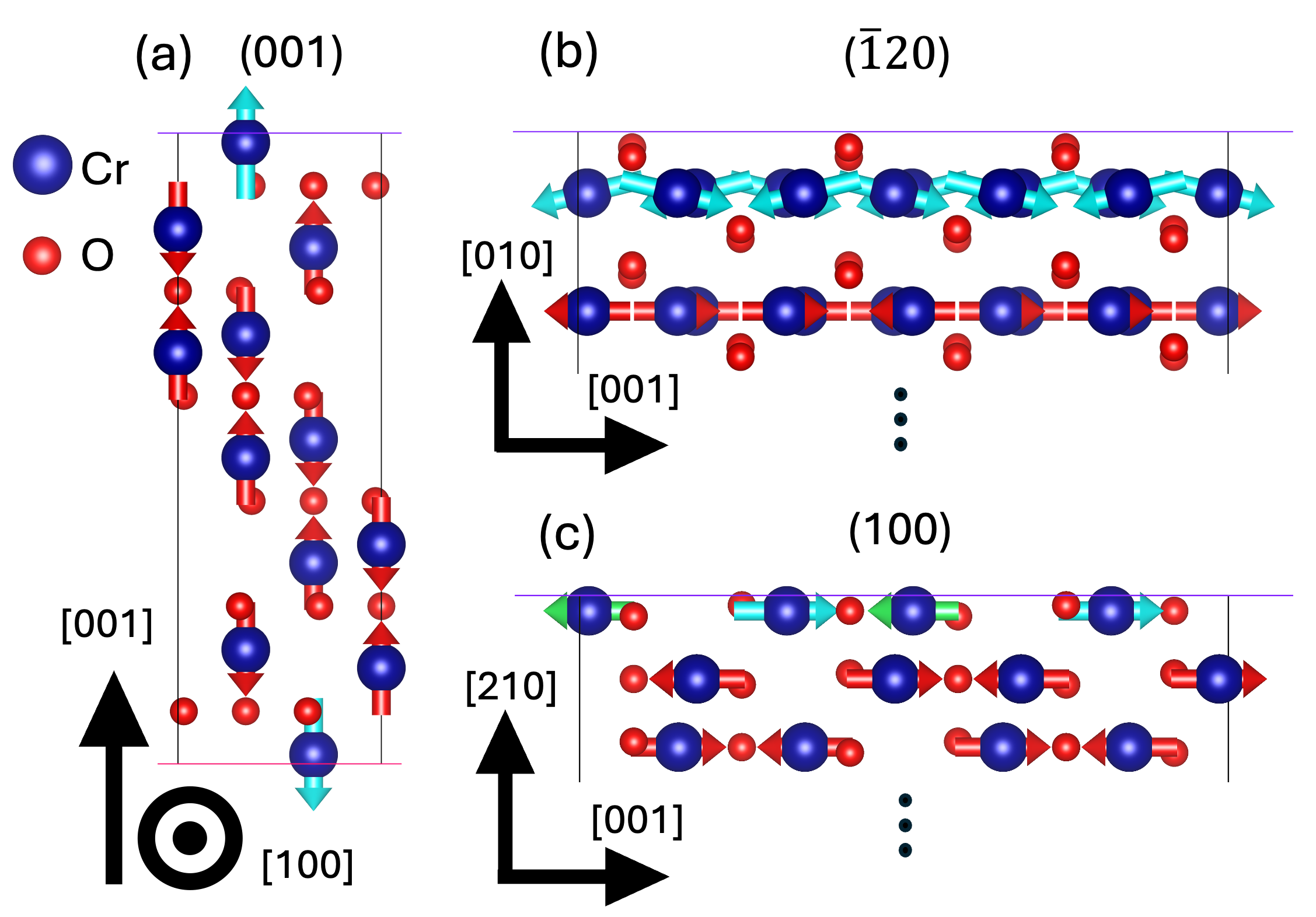}
    \caption{Unit cells for the (a) (001), (b) ($\mathrm{\bar{1}20}$), and (c) (100) slab geometries of $\mathrm{Cr_2O_3}$. (b) and (c) show the first few layers near the surface. The red arrows show the $\mathrm{Cr}$ bulk magnetic moments. The blue/green arrows are magnetic moments of $\mathrm{Cr}$ surface atoms which generate the surface magnetization. For the surfaces perpendicular to (001), the real-space surface normals and their corresponding Miller planes do not have the same indices.}
    \label{fig:Cr2O3Lattices}
\end{figure}

\textit{Ab-initio results} -- We start with bulk AFM $\mathrm{Cr_2O_3}$, which crystallizes in the corundum structure with magnetic space group (point group) $R\text{-}3'c'$ ($-3'$). $\mathrm{Cr_2O_3}$ is a linear magnetoelectric~\cite{E_011_03_0708, Dzyaloshinski1960OnTM}, meaning it acquires an induced magnetization in response to applied electric field and conversely, a net electric polarization in response to an applied magnetic field. The room-temperature ordering and efficient switching of its Néel vector make it attractive for spintronics applications\cite{schlitzEvolutionSpinHall2018,kosubAllElectricAccessMagneticFieldInvariant2015}. Fig.~\ref{fig:Cr2O3Lattices} shows the three $\mathrm{Cr_2O_3}$ surface facets $(001)$, $(100)$, and $(\bar{1}20)$ of interest in this Letter, for which net surface magnetization has been confirmed experimentally and theoretically~\cite{wuImagingControlSurface2011, appelNanomagnetismMagnetoelectricGranular2019, wornleCoexistenceBlochNeel2021,heRobustIsothermalElectric2010, borisovMagnetoelectricSwitchingExchange2005}. $(001)$ $\mathrm{Cr_2O_3}$ (Fig.~\ref{fig:Cr2O3Lattices}(a)) is a magnetically uncompensated surface with an unpaired Cr sublattice. In contrast, the $(\bar{1}20)$ and $(100)$ crystallographic planes (Figs.~\ref{fig:Cr2O3Lattices}(b) and (c)), which are parallel to the bulk Néel vector, have perfectly compensated AFM order in the bulk, but acquire an induced magnetization at surfaces and interfaces.

\begin{figure}
    \centering
    \includegraphics[width=3.4in]{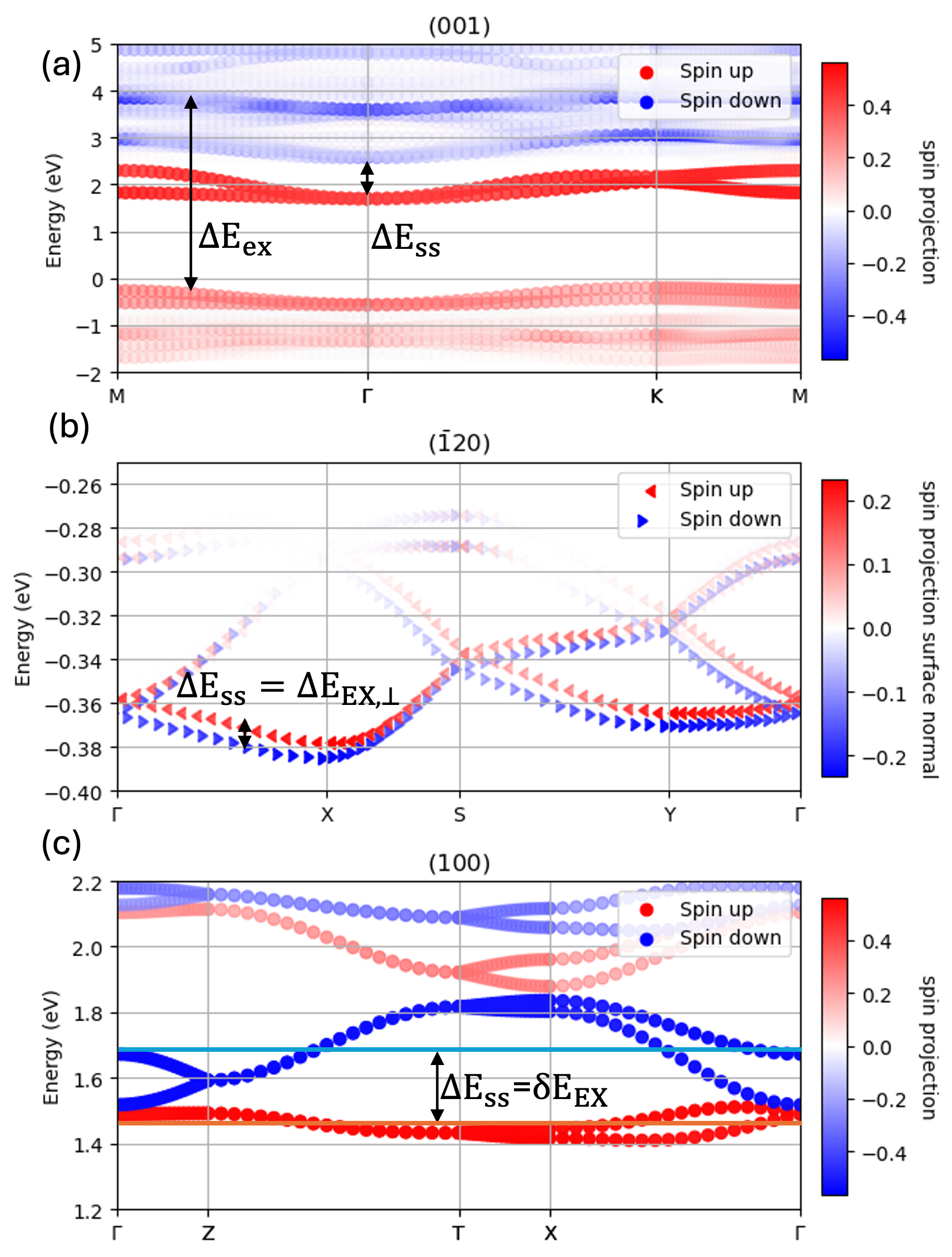}
    \caption{Surface- and spin- projected  band structures for the (a) (001), (b) ($\bar{1}20$) and (c) (100) surfaces of $\mathrm{Cr_2O_3}$.}
    \label{fig:Cr2O3bands}
\end{figure}

We first consider the $(001)$ surface of $\mathrm{Cr_2O_3}$ (Fig.~\ref{fig:Cr2O3Lattices}(a)), whose magnetic point group $3$ permits a net magnetization along the [001] surface normal, aligned with the bulk Néel vector. Fig.~\ref{fig:Cr2O3bands}(a) shows the projected band structure of this surface, with spin-orbit coupling (SOC) neglected (SM, see Appendix A~\cite{SM}, for computational details).
We see a clear spin polarization due to the onsite exchange splitting, $\Delta E_{EX}\approx4\mathrm{eV}$ for the labeled representative surface Cr $d$ band (Fig.~\ref{fig:Cr2O3bands}(a)), consistent with the single-sublattice model of Eq.~\ref{Hamiltonian,on-site}. We identify pairs of exchange-split bands by comparing orbital projections, as detailed in SM, Appendix C~\cite{SM}. The corresponding spin splitting for this specific surface is smaller, around $0.9~\mathrm{eV}$ at the $\Gamma$ point, due to the superposition of other $d$ bands in the same energy window.

We next consider the nominally compensated $(\bar{1}20)$ surface of $\mathrm{Cr_2O_3}$, with magnetic point group $2$~\cite{weberSurfaceMagnetizationAntiferromagnets2024}. This point group has a two-fold rotation about the surface normal which connects opposite $\mathrm{Cr}$ sublattices, so surface magnetization arises exclusively from canting along the surface normal, perpendicular to the in-plane Néel vector (Fig.~\ref{fig:Cr2O3Lattices}(b)). The lowest-energy surface magnetic state corresponds to an out-of-plane canting angle of approximately $\pm 0.25^{\circ}$, with the sign dependent on the sign of the bulk Néel vector~\cite{weberSurfaceMagnetizationAntiferromagnets2024}.

Fig. ~\ref{fig:Cr2O3bands}(b) shows the surface-projected band structure of a $(\bar{1}20)$ slab, where we have initialized the energetically favorable $0.25^{\circ}$ canting of surface $\mathrm{Cr}$ moments and included SOC. We project the bands onto spin components along the canting direction, perpendicular to the in-plane Néel vector, and denote the corresponding projected exchange splitting as $\Delta E_{EX,\perp}$.

The average spin splitting for the representative $d$ band close to the Fermi level (Fig.~\ref{fig:Cr2O3bands}(b)) is around 0.01 eV. The small size of this relativistic spin splitting $\Delta E_{ss}=\Delta E_{EX,\perp}$, compared to the $\sim 1\mathrm{eV}$ spin splitting at the $(001)$ surface, is consistent with our discussion in the previous section (Eqs.~\ref{Hamiltonian,on-site}-\ref{Hamiltonian-120} and Fig.~\ref{fig:splitting}(b)). 

The last $\mathrm{Cr_2O_3}$ surface orientation that we consider is $(100)$ (Fig.~\ref{fig:Cr2O3Lattices}(c)), which has magnetic point group $m'$. This point group allows a net magnetization along [001], with a magnitude of $\sim~0.05\mu_B/\mathrm{nm^2}$ based on our present DFT calculations and prior work\cite{weberSurfaceMagnetizationAntiferromagnets2024}. Although $m'$ also permits magnetization perpendicular to the surface via canting\cite{pylypovskyiSurfaceSymmetryDrivenDzyaloshinskiiMoriyaInteraction2024}, here we constrain surface $\mathrm{Cr}$ moments to remain collinear to isolate the $[001]$ contribution. Including the relativistic out-of-plane magnetization component affects the band structure negligibly (see SM, Appendix G~\cite{SM}).

The mirror plane in $m'$ connects Cr belonging to the \textit{same} sublattice; no surface symmetry connects opposite sublattices. Thus, unlike for the $(\bar{1}20)$ surface, opposite-spin Cr sublattices at $(100)$ surface are symmetry-allowed to have differing crystal field environments. However, for the electrostatically stable termination of the $(100)$ surface, the oxygen coordination of the two sublattices is very similar, so $\Delta E_{CF}$ is negligible (see SM, Appendix E~\cite{SM}) and the spin splitting is dominated by the second term in Eq.~\ref{SpecificHamiltonian} (Fig.~\ref{fig:splitting}(g)).

In SM, Appendix C~\cite{SM}, we show that the two $(100)$ surface sublattices have similar but inequivalent on-site splittings close to $3~ \mathrm{eV}$ for the $d$ bands considered in Fig.~\ref{fig:Cr2O3bands}(c). Fig.~\ref{fig:Cr2O3bands}(c) shows the surface-projected band structure with contributions from both sublattices. A small spin splitting of $\sim 0.2\ \mathrm{eV}$ is present, corresponding to the $\delta E_{EX}$ term in Eq.~\ref{SpecificHamiltonian}. As expected from Fig.~\ref{fig:splitting}(f)-(g), the splitting is small compared to the on-site exchange splittings, since opposite-sign exchange splittings of the two sublattices partially cancel each other in this energy range.

\begin{figure}
    \centering
    \includegraphics[width=3.4in]{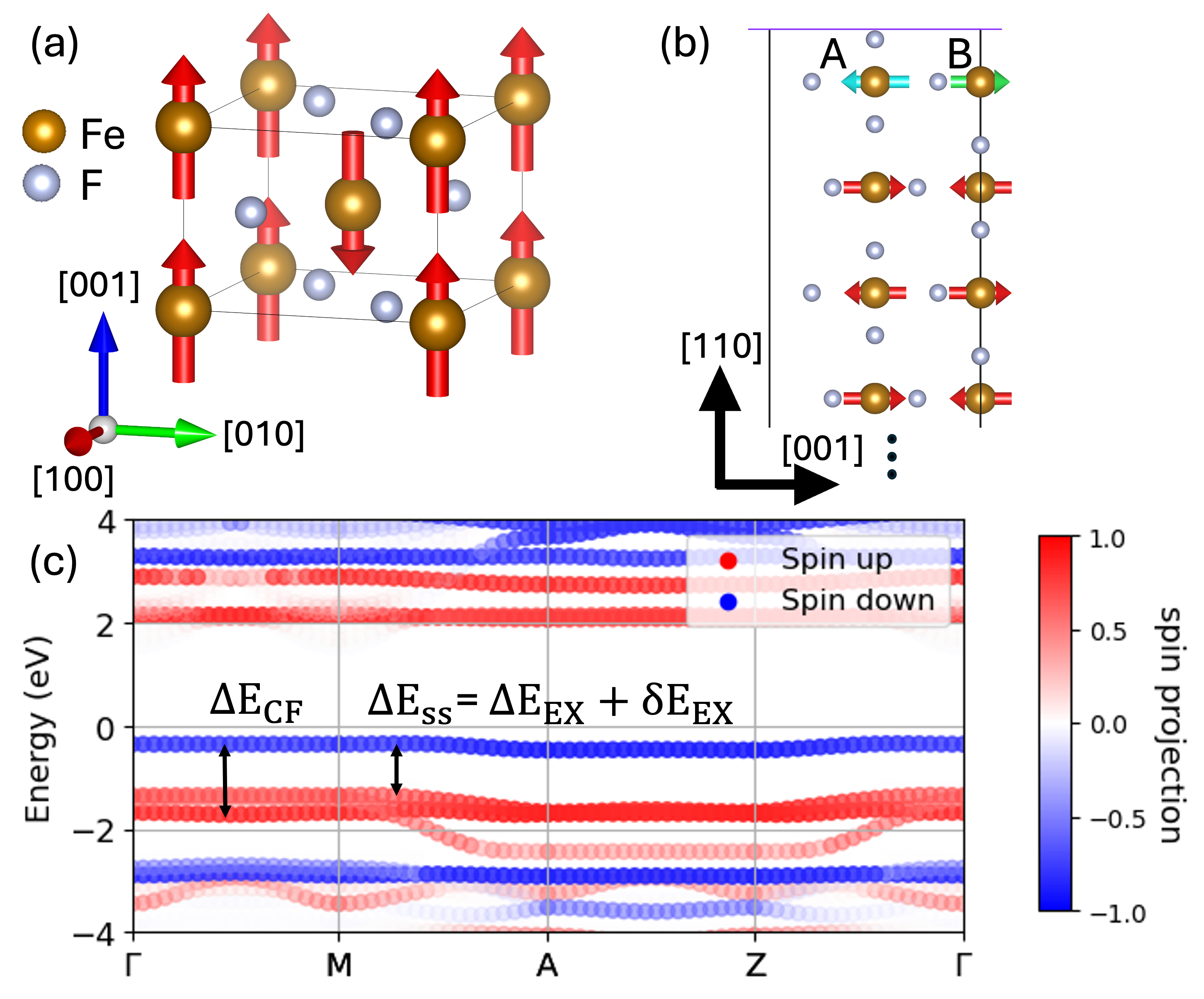}
    \caption{(a) Bulk structure of $\mathrm{FeF_2}$. (b) First few layers near the (110) surface for the slab used in our DFT calculations. The blue/green arrows represent the inequivalent surface magnetic moments while the red arrows show the bulk-like moments deeper in the slab. (c) The spin- and $(110)$ surface-projected band structure.}
    \label{fig:FeF2bands}
\end{figure}

The final surface we consider belongs to $\mathrm{FeF_2}$, a bulk altermagnet~\cite{smejkalEmergingResearchLandscape2022} with magnetic space group (point group) $P4_2'/mnm'$ ($4/mmm$). Fig.~\ref{fig:FeF2bands}(a) shows the bulk crystal structure. We examine the (110) surface, which contains two oppositely pointing Fe sublattices (Fig.~\ref{fig:FeF2bands} (b)). The surface magnetic point group is $m'm2'$, which allows a net $~\sim0.05\mu_B/\mathrm{nm^2}$ magnetization exclusively along the in-plane [001] Néel vector direction\cite{weberSurfaceMagnetizationAntiferromagnets2024}, via an inequivalence in the sublattice moment magnitudes (Fig.~\ref{fig:FeF2bands} (b)). No symmetries in $m'm2'$ connect the Fe sublattices, so this surface has the same motif as $(100)$ $\mathrm{Cr_2O_3}$, described by Eq.~\ref{SpecificHamiltonian}. In contrast to $(100)$ $\mathrm{Cr_2O_3}$ however, where crystal fields of opposite Cr sublattices are nearly identical, for the electrostatically stable termination of $(110)$ $\mathrm{FeF_2}$, one Fe sublattice (labeled B) is only five-fold coordinated by F atoms, whereas the other Fe sublattice (label A) retains its bulk six-fold coordination (Fig.~\ref{fig:FeF2bands}(b)). Thus, we expect the Fe $d$ state energies set by the CF environments to differ drastically, making the $\Delta E_{CF}$ term in Eq.~\ref{SpecificHamiltonian} appreciable.

Fig.~\ref{fig:FeF2bands}(c) shows the total surface-projected bandstructure of $(110)$ $\mathrm{FeF_2}$, with a large spin splitting of $1~\mathrm{eV}$. Based on our sublattice-projected band structure (see SM, Appendix D~\cite{SM}) $\Delta E_{ss}$ is precisely the on-site exchange splitting $\Delta E_{EX}+\delta E_{EX}$ for the six-fold-coordinated sublattice band pair. Crucially, the large $\Delta E_{CF}\approx1.3~\mathrm{eV}$ energetically separates sublattice bands, yielding a net splitting an order of magnitude larger than the spin splitting on $(100)$ $\mathrm{Cr_2O_3}$ despite nearly identical values of net surface magnetization. The spin splitting magnitude of $(110)$ $\mathrm{FeF_2}$ can also be contrasted with the maximum value $\sim0.08~\mathrm{eV}$ of its bulk altermagnetic spin splitting in the valence band~\cite{zouGiantTunnelingMagnetoresistance2026}.

\textit{Conclusion} -- We have demonstrated that ferromagnetic surfaces of AFMs provide a general platform for large spin splitting, on the scale of conventional FMs and the largest reported values for altermagnets. While exchange associated with the net surface magnetization is \textit{necessary} for spin-split bands, its magnitude can be strongly enhanced by spin-independent energy scales. In particular, crystal-field terms set by the surface termination can separate bands belonging to opposite magnetic sublattices, producing large spin splitting even when the net surface magnetization is small, as in $(110)$ $\mathrm{FeF_2}$.

Because low-symmetry terminations leading to net surface magnetization are common, our findings substantially expand the AFM materials space for functional spin splitting. Moreover, this mechanism for AFM spin splitting has the potential to unlock new applications as it enables normally contraindicated properties to coexist in a single material, for example spin splitting at $\mathcal{I}\mathcal{T}$-broken surfaces together with magnetoelectricity in an $\mathcal{I}\mathcal{T}$-preserving bulk such as $\mathrm{Cr_2O_3}$.\\
\indent Although the bulk AFMs and surfaces studied here are insulating, the same symmetry arguments apply to metallic surfaces. In that case, spin-split surface bands can generate time-reversal-broken transport responses, such as an anomalous Hall effect, providing a route to bulk Néel-vector readout~\cite{taoLayerHallDetection2024}. We note that surface-derived transport for Néel vector detection requires decoupling opposite-surface contributions when they have opposite-sign spin splitting, as in bulk $\mathcal{I}\mathcal{T}$-symmetric $\mathrm{Cr_2O_3}$. This becomes possible when the bulk is insulating while surface bands cross the Fermi level through intrinsic surface-state energy shifts~\cite{zhaoEmergentAnomalousHall2026} or selective doping. In $\mathrm{FeF_2}$, where opposite surfaces have the same sign of spin splitting, similar constraints on the bulk conductivity are absent (see SM, Appendix F~\cite{SM}).
\indent Finally, our results are timely given recent proposals of surface altermagnetism \cite{jaeschke-ubiergoEmergentAltermagnetismSurfaces2026}. The exchange-driven mechanism identified here arises generically from finite surface magnetization and yields large spin splitting at surfaces with an uncompensated sublattice, 
\textit{or} opposing sublattices that are not symmetry-related: neither of these surface motifs support altermagnetism. Our work highlights how these low-symmetry terminations unlock additional energy scales on surfaces complementary to those hosting altermagnetism.\\
\indent Overall, our findings reveal a new platform for large, functional spin splitting in AFMs, and more broadly, highlight how the intrinsic symmetry-lowering of surfaces can enable functional properties forbidden in the bulk.
\newline

\textbf{Acknowledgments}\\ 
The authors would like to thank Nicola Spaldin, Denys Makarov and Raihan Ahammed for useful discussions. WAS and SFW acknowledge funding from Chalmers University of Technology through the department of Physics and Astronomy. SFW also acknowledges funding through the Areas of Advance Nano and Materials Science. DFT simulations were performed using computing resources from the National Supercomputer Centre (NSC) at Linköping University. These resources were granted by the National Academic Infrastructure for Supercomputing in Sweden (NAISS), partially funded by the Swedish Research Council through grant agreement no. 2022-06725. 

\section*{Appendix}
\subsection{DFT calculation details for $\mathrm{Cr_2O_3}$ $(001)$, $(\bar{1}20)$ and $(100)$ surfaces} \label{DFTCr2O3}
For our density functional theory (DFT) calculations we use the Vienna Ab initio Simulation package (VASP)  \cite{kresseEfficientIterativeSchemes1996}. For the ($\mathrm{\bar{1}20}$) slab geometry, the noncollinear local spin-density approximation (LSDA) is used with spin-orbit coupling (SOC) included self-consistently. In the case of the (001) and (100) slab geometries, SOC is omitted and instead we use collinear spin-polarized calculations. We adopt the projector augmented wave method (PAW) with the standard VASP PAW pseudopotentials \cite{blochlProjectorAugmentedwaveMethod1994}. These have the following valence electron configurations: Cr $3p^6 4s^1 3d^5$ and O $2s^2 2p^4$. The default integration spheres are used, which are 1.164 \AA\ and 0.820 \AA\ respectively. In line with previous calculations \cite{weberSurfaceMagnetizationAntiferromagnets2024, weberCharacterizingOvercomingSurface2023} we apply a Hubbard correction of 4 eV for the $d$ orbitals for Cr \cite{anisimovFirstprinciplesCalculationsElectronic1997}. Gaussian smearing is chosen with a smearing width of 0.02 eV. For all calculations, we use a kinetic energy cutoff of 800 eV.

To model the (001) surface, we use a hexagonal unit cell of $\mathrm{Cr_2O_3}$ with 12 $\mathrm{Cr}$ atoms. We first relax the bulk structure until forces on all atoms are less than 0.01 eV/\AA. This leads to an in-plane lattice vector of 4.91 \AA\ and an out-of-plane lattice vector along $[001]$ of 13.52 \AA. To study the $(001)$ surface we create a slab structure from this relaxed bulk cell by adding 20 \AA\ of vacuum in the [001] direction. We then relax the atomic positions until the forces on the atoms are less than 0.01 eV/\AA\ and the stresses are below a stopping criterion which is internally scaled in VASP based on the selected force value. We use a $\Gamma$-centered k-point mesh of $9\times 9 \times 5$ for the bulk structure and $9\times 9 \times 5$ for the (001)-oriented vacuum-terminated slab.

To model the (100) surface, we construct an orthorhombic unit cell, starting from the relaxed hexagonal bulk unit cell. This unit cell has in-plane orthogonal lattice parameters of lengths 4.91 and 13.52 \AA. The smallest electronically stable unit cell is copied 4 times in the direction perpendicular to the $(100)$ surface such that a 12-layer structure is created, and 15 \AA\ of vacuum is added along the $(100)$ surface normal. The final structure has 48 Cr and 72 O atoms in total. The atomic positions are again relaxed while keeping the lattice vectors constant. We use a $\Gamma$-centered k-point mesh of $8\times 3 \times 1$.

Finally, to model the ($\mathrm{\bar{1}20}$) surface of $\mathrm{Cr_2O_3}$ we also use an orthorhombic unit cell with orthogonal in-plane lattice vectors of 8.51 and 13.52 \AA. The smallest electronically stable unit cell is copied two times to get a 4-layer structure with 48 Cr atoms and 72 O atoms. To this structure, 15 \AA\ of vacuum is added along the surface normal and atomic positions are again relaxed using the same force and stress criteria as above. We use a $\Gamma$-centered k-point mesh of $8\times 5 \times 1$.

To model the finite surface magnetization arising from out-of-plane canting for the $(\bar{1}20)$ surface case, we use constrained magnetic calculations~\cite{maConstrainedDensityFunctional2015} to fix the $\mathrm{Cr}$ magnetic moments in the central two layers along their bulk in-plane $[001]$ direction. We cant the moments in the top and bottom layers $0.25^{\circ}$ towards bulk, based on the energetic minimum as a function of canting angle calculated already in Ref.~\cite{weberSurfaceMagnetizationAntiferromagnets2024}. We use a Lagrange multiplier of $\lambda=10$, which is sufficient to fix the magnetic moments while keeping the added penalty energy negligible compared to the total energy of the system.

\subsection{DFT calculation details for $\mathrm{FeF_2}$ (110)} \label{DFTFeF2}
For calculations of the $\mathrm{FeF_2}$ surface band structure, we also use the VASP software package. We use the Perdew-Burke-Ernzerhof (PBE) functional \cite{perdewGeneralizedGradientApproximation1996} and select a Hubbard U of 6 eV and Hund's exchange J of 0.95 eV, based on previous work \cite{weberSurfaceMagnetizationAntiferromagnets2024, lopez-morenoFirstprinciplesStudyElectronic2012, munozSurfaceStatesFeF22015}. For the PAW pseudopotentials used, Fe and F have the following valence electron configurations: $3d^7 4s^1$ and $2s^2 2p^5$. Since the surface magnetization which develops for $(110)$ $\mathrm{FeF_2}$ is fully collinear along the direction of the bulk Néel vector, we neglect SOC and use collinear spin-polarized DFT. Gaussian smearing is selected with a smearing width of 0.02 eV.

We relax the bulk $\mathrm{FeF_2}$ structure using a cutoff energy of 800 eV for the kinetic energy and a $6\times 6 \times 9$ Gamma-centered k-point mesh for the bulk tetragonal cell. This leads to relaxed lattice vectors of lengths 4.78 and 3.34 \AA. For the (110) surface, we construct a tetragonal unit cell with the in-plane lattice vectors of lengths 6.76 and 3.34 \AA. The smallest electronically stable unit cell is extended along the [110] direction to 8 layers and 15 \AA\ of vacuum is added in the [110] direction. The atomic positions are relaxed  while keeping the lattice vectors fixed with the same force and stress criteria as for $\mathrm{Cr_2O_3}$. We use a $6\times 9 \times 1$ Gamma-centered k-mesh for the slab relaxation.

\subsection{Exchange splitting in $\mathrm{Cr_2O_3}$} \label{AppendixEXCr2O3}
The surface-projected band structures of $\mathrm{Cr_2O_3}$ and $\mathrm{FeF_2}$ can of course not be mapped exactly to our toy models (Eqs.\ref{Hamiltonian,on-site}-\ref{SpecificHamiltonian} in the main text) containing only one orbital per magnetic sublattice. Even when we focus on splitting of specific representative bands close to the Fermi level, as we have done in the main text, complexity of the band structure, in particular orbital hybridization with bulk bands at similar energy levels, makes unambiguous assignment of specific pairs of on-site exchange-split bands difficult in some cases. We make educated guesses by assuming that the spin-up and spin-down bands of an exchange-split pair will have similar orbital character, though we emphasize again this is not always the case if hybridization with nearby bulk bands occurs. 

\begin{figure*}
    \centering
    \includegraphics[width=6.8in]{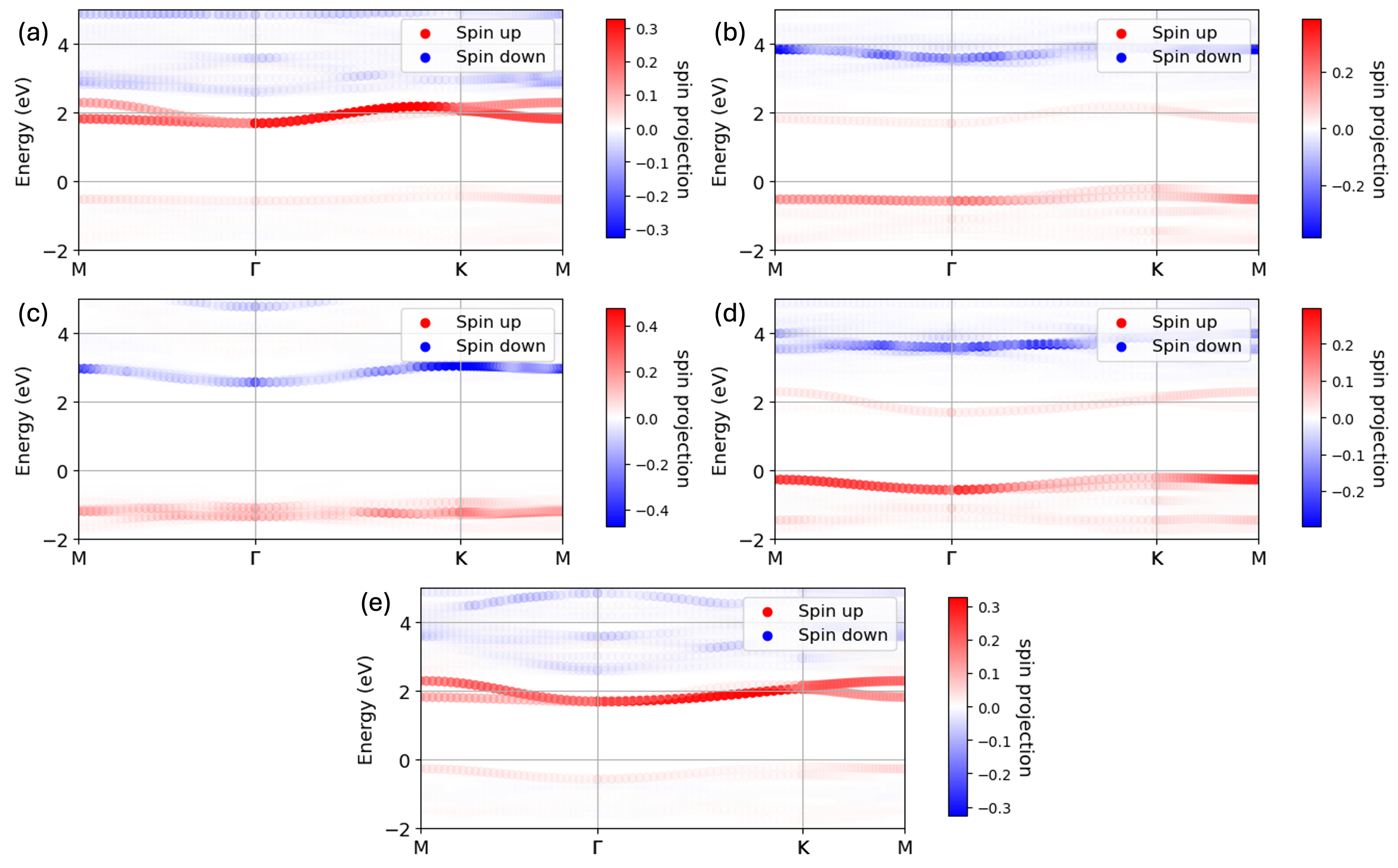}
    \caption{Projection of the spin of specific d orbitals on the surface $\mathrm{Cr_2O_3}$ atom. We show (a) $\mathrm{d_{xy}}$, (b) $\mathrm{d_{yz}}$, (c) $\mathrm{d_{z^2}}$, (d) $\mathrm{d_{xz}}$ and (e) $\mathrm{d_{x^2-y^2}}$}.
    \label{fig:all d orbitals Cr2O3}
\end{figure*}

For the (001) surface of $\mathrm{Cr_2O_3}$ for example, in Figs.~\ref{fig:all d orbitals Cr2O3} (a)-(e) we plot the projections of the five different $d$ orbitals onto bands in the energy range shown in Fig.~\ref{fig:Cr2O3bands}(a) in the main text. In this region both the $d_{yz}$ and $d_{xz}$ bands have a large projection. Of these we highlight the pair of spin-up and spin-down bands with a strong $d_{xz}$ projection. Based on this, we assign an on-site exchange splitting of $\approx 4\mathrm{eV}$ for the highest spin-up valence band belonging to the uncompensated $\mathrm{Cr}$ sublattice on the $(001)$ surface. This is also labeled in Fig.~\ref{fig:Cr2O3bands}(a) in the main text. We note as well that the relative position of the spin polarized bands (i.e., spin up occupied and spin-down unoccupied) is consistent with the positive, outwards magnetization of the surface Cr atom in our DFT calculations.

\begin{figure}
    \centering
    \includegraphics[width=3.4in]{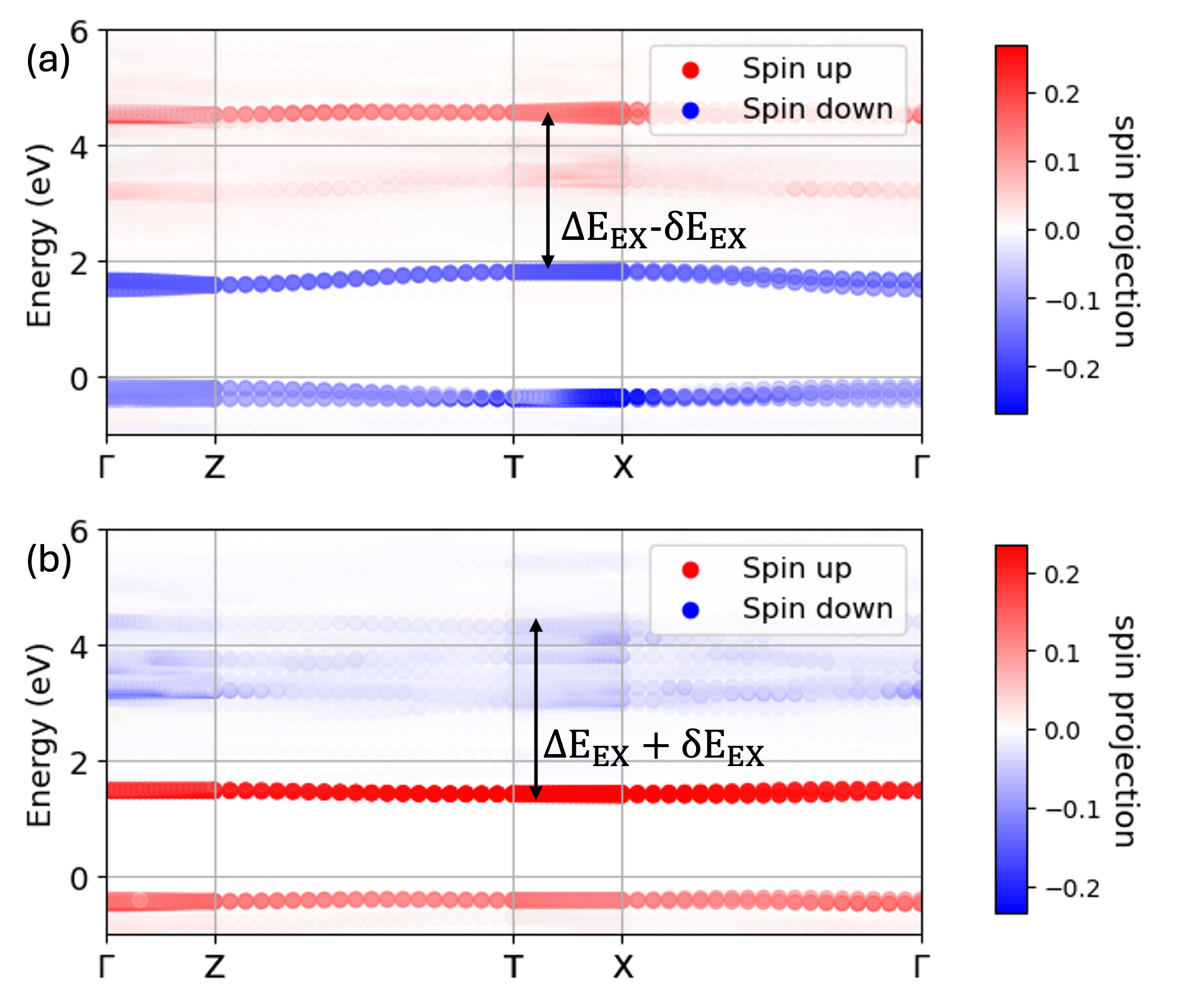}
    \caption{Projection of the spin on the $d_{xy}$ orbital of the magnetic sublattice with the smaller (a) and larger (b) net magnetization at the (100) surface.}
    \label{fig:dxy orbital Cr2O3 100}
\end{figure}

For the (100) surface of $\mathrm{Cr_2O_3}$, we can do a similar analysis. Here, in Figs.~\ref{fig:dxy orbital Cr2O3 100}(a)-(b) we plot separately projections onto the two oppositely pointed Cr sublattices at the (100) surfaces. Based on similar projections of $d_{xy}$ orbitals, we identify the labeled pairs in each cases as likely related through onsite exchange splittings of $\Delta E_{EX}+\delta E_{EX}$ and $\Delta E_{EX}-\delta E_{EX}$ respectively. From Figs.~\ref{fig:dxy orbital Cr2O3 100}(a)-(b) it is also clear that the crystal field inequivalence $\Delta E_{CF}$ is negligible, since the centers of the sublattice-projected exchange-split band pairs are roughly the same. Thus, the net spin polarization near the Fermi level for the the (100) surface, as shown in Fig.~\ref{fig:Cr2O3bands}(c) in the main text, is on the order of the small difference between on-site exchange splittings, $\delta E_{ex}\approx 0.2 \mathrm{eV}$.

\subsection{Exchange splitting and Crystal Field inequivalence in $\mathrm{FeF_2}$} \label{AppendixEXFeF2}
\begin{figure}
    \centering
    \includegraphics[width=3.4in]{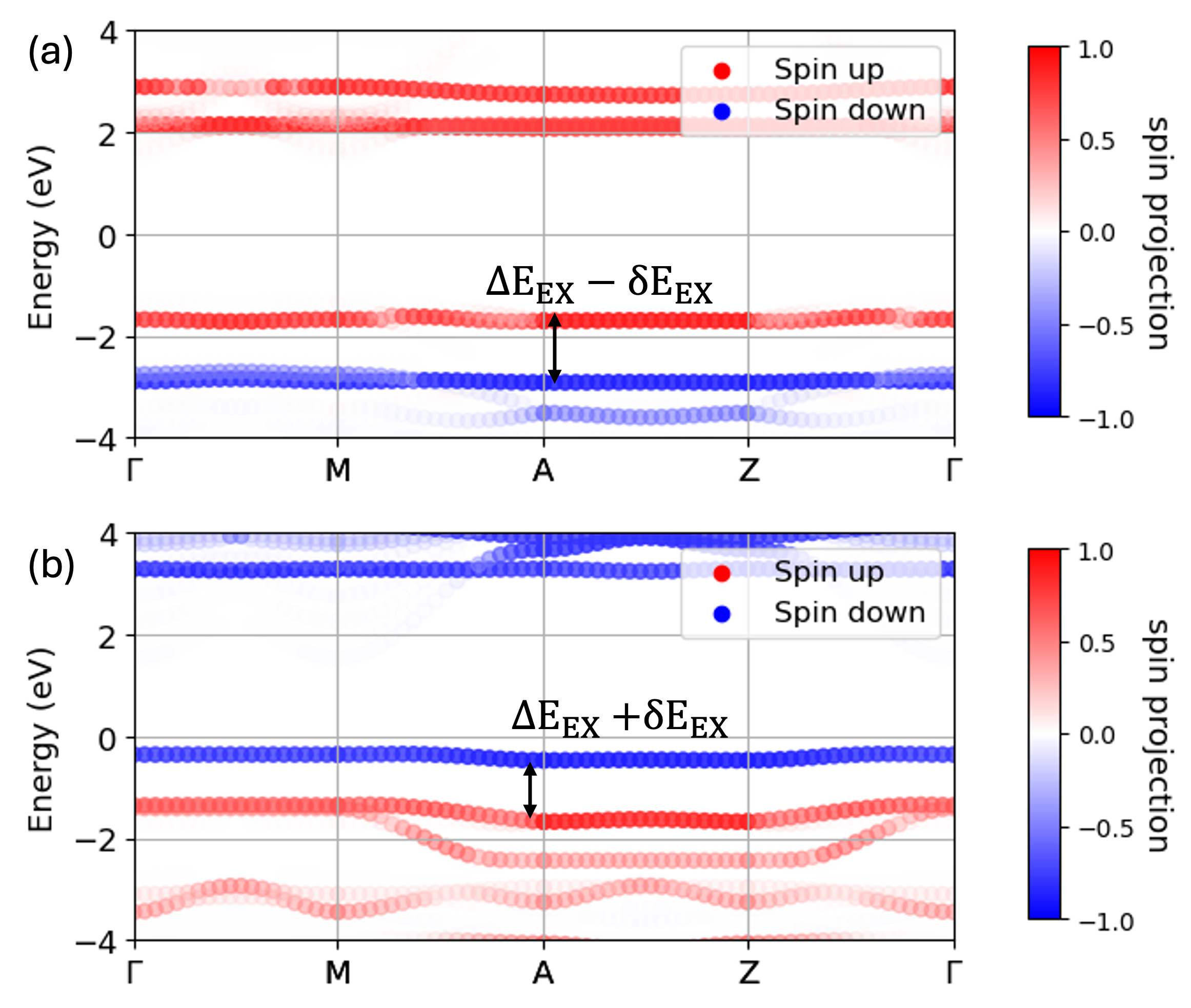}
    \caption{Spin projection of the $d$ bands for (a) the 5-fold surface sublattice and (b) 6-fold surface sublattice.}
    \label{fig:FeF2sublattices}
\end{figure}

\begin{figure}
    \centering
    \includegraphics[width=3.4in]{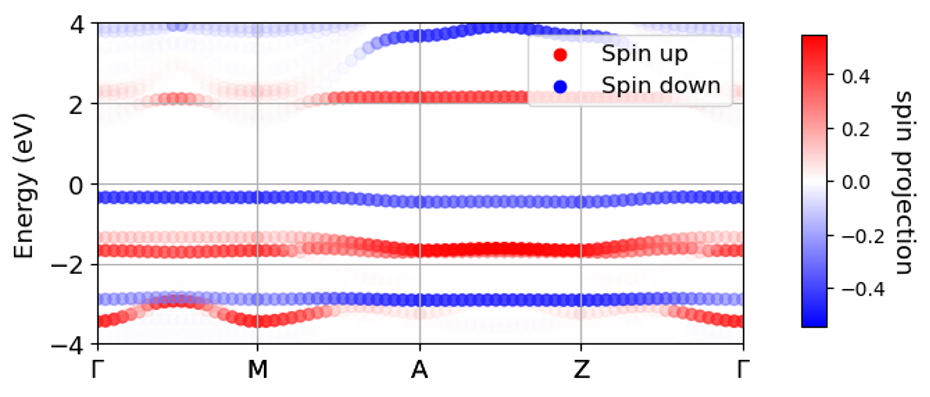}
    \caption{Projection of the spin along the Néel vector on the $d_{xy}$ orbital of the Fe sites at the (110) surface.}
    \label{fig:dxy orbital FeF2}
\end{figure}

We first justify our claim in the main text that the quantitative (110) surface spin polarization of $\mathrm{FeF_2}$ near the Fermi level is primarily determined by the sublattice crystal field inequivalence $\Delta E_{EX}$. In Fig.~\ref{fig:FeF2sublattices}(a)-(b) we show the $(110)$ projected band structure of $\mathrm{FeF_2}$ projected separately onto the two oppositely-pointed Fe sublattices which combine into Fig.~\ref{fig:FeF2bands}(c). We see that the highlighted groups of $d$ bands have opposite spin polarization when comparing the two sublattice projections, as expected, but additionally, the $d$ manifold for the 5-fold coordinated $\mathrm{Fe}$ sublattice is shifted rigidly by about $\Delta E_{CF}\approx$ 1.3 eV compared to that for the 6-fold coordinated sublattice. This crystal field-mediated shift is highlighted in Fig. \ref{fig:FeF2bands} of the main text.  
The $d_{xy}$ orbital is given in Fig. \ref{fig:dxy orbital FeF2}. Here we see a clear gap of around 1 eV between two spin-polarized bands. This is quite a bit smaller than what we found for $\mathrm{Cr_2O_3}$.

As explained in the main text, the large $\Delta E_{CF}$ term for (110) $\mathrm{FeF_2}$ energetically separates the on-site exchange-split bands for the two $\mathrm{FeF_2}$ sublattices such that the bands no longer overlap. We perform a similar-orbital projected analysis to identify likely on-site exchange-split pairs for the $(110)$ surface of $\mathrm{FeF_2}$. Based on similar $d_{xy}$ projections for the labeled bands in Fig. \ref{fig:dxy orbital FeF2}, we identify inequivalent on-site exchange splittings of $\approx$1.2 (1) eV for the five- (six-) fold coordinated Fe sublattices for the $d$ bands of interest. When these two sublattice-projected band structures are combined, the overall spin polarization of the surface band structure is still proportional to the on-site exchange energies due to the crystal-field mediated separation of sublattice bands. This is in contrast to (100) $\mathrm{Cr_2O_3}$, there where opposite, slightly inequivalent exchange splittings of the two sublattice overlap, leading to a reduction in net spin polarization.

\subsection{Oxygen coordination $\mathrm{Cr_2O_3}$ $(100)$} \label{AppendixOxygenCoordinationCr2O3}
\begin{figure}
    \centering
    \includegraphics[width=3.4in]{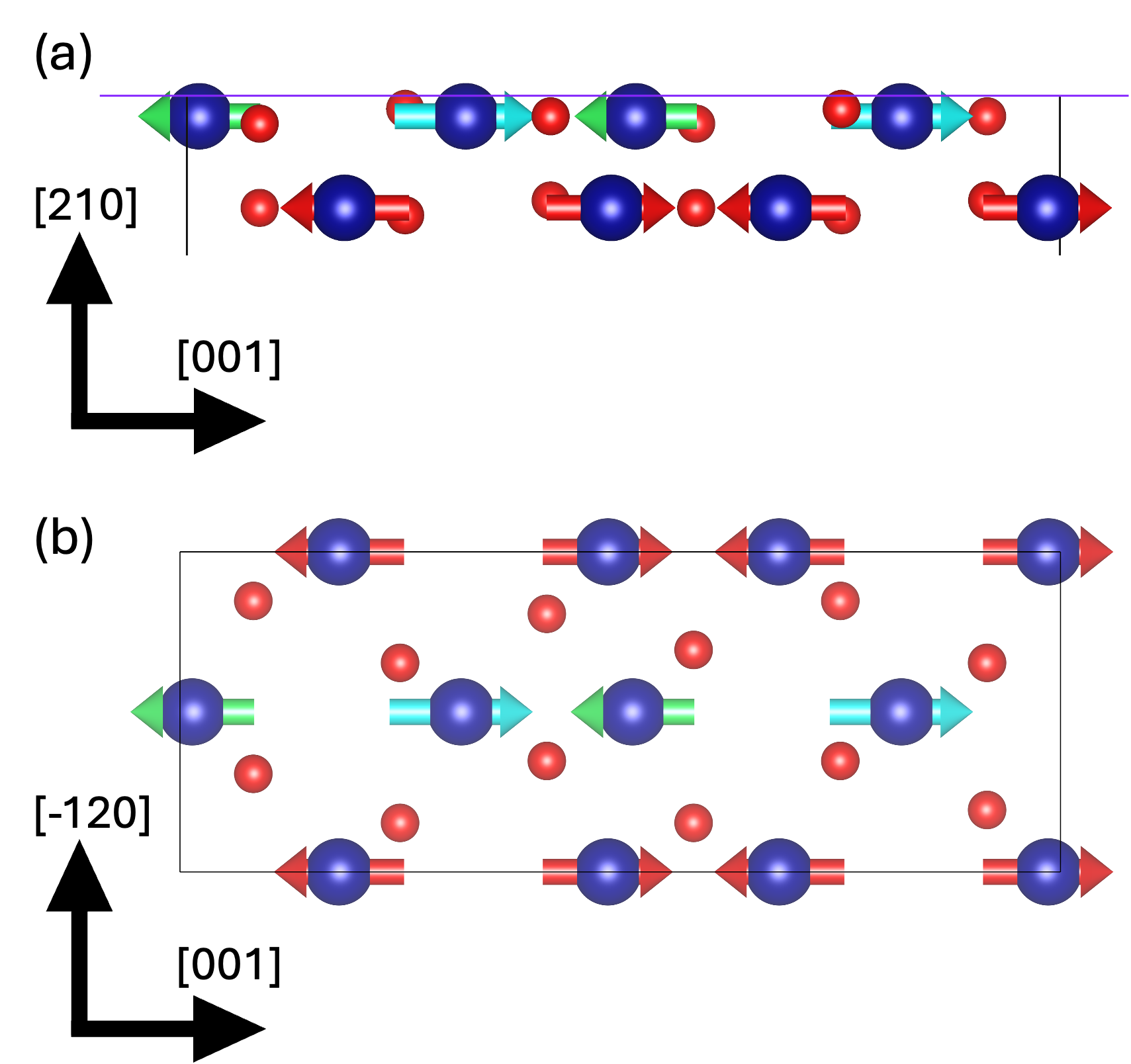}
    \caption{(a) side view and (b) top view of the unit cell for $\mathrm{Cr_2O_3}$ (100) slab geometry. Only the top two layers are given in both figures.}
    \label{fig:Cr2O3lattice}
\end{figure}

At the (100) surface of $\mathrm{Cr_2O_3}$ the magnetic sublattices are not connected by a symmetry. As a result a net magnetization is allowed along the Néel vector. However this same lack of connecting symmetry allows for differences in the ligand crystal fields for the two magnetic sublattices. However, in contrast to the case of $(110)$ $\mathrm{FeF_2}$, in which removal of one of the surface F atoms for the electrostatically stable termination leads to vastly different crystal field energies for different Fe sublattices, the coordination of oxygen atoms around the (100) surface chromium atoms is very similar for the electrostatically stable termination. This can be seen in Fig.~\ref{fig:Cr2O3lattice}, from both the side view (a) and the top view (b). Notably, oxygen atoms in the same layer as the surface Cr are identically positioned around the two sublattices. A slight difference in $d$ state energies can arise due to the inequivalent distances of the oxygen atoms in the first \textit{subsurface} layer to the surface Cr sublattice, which can be seen in (Fig.~\ref{fig:Cr2O3lattice}(a)). However, the relative shift in energies of sublattice-projected Cr $d$ states as a result of this coordination distance for next-nearest oxygen neighbors is negligible (as evident in the sublattice-projected band structure in Appendix \ref{AppendixEXCr2O3}.) Hence unlike the case for $(110)$ $\mathrm{FeF_2}$, the overall spin-split band structure in (\ref{fig:Cr2O3bands}(c)) of the main text is dictated primarily by the small exchange splitting $\delta E_{EX}$ due to net surface magnetization, and not by $\Delta E_{CF}$.

\subsection{Symmetry relation between surfaces of slab} \label{AppendixOppositeSurfaces}

\begin{figure}
    \centering
    \includegraphics[width=3.4in]{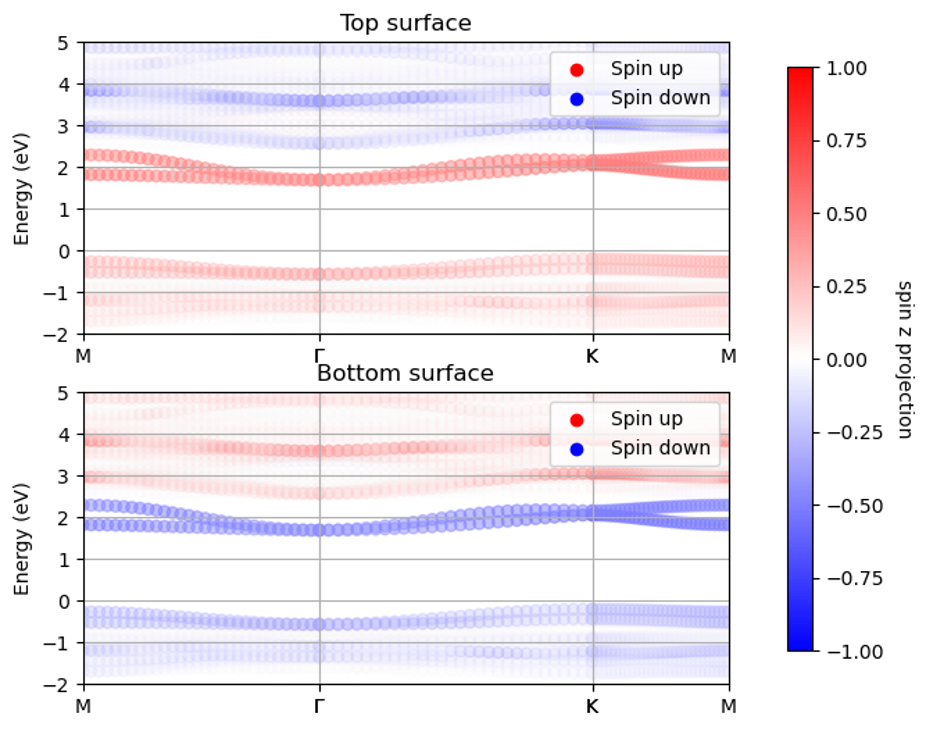}
    \caption{Spin projected at the top/bottom surface for a (001) slab geometry of $\mathrm{Cr_2O_3}$. The spin projection is opposite comparing the bottom and top surface. The surfaces consist of 1 Cr site each.}
    \label{fig:Cr2O3topbot}
\end{figure}

For experimental verification and probing of the surface-localized spin splitting discussed in this work, and the associated transport effects, it is important to consider how the spin splitting of opposite surfaces are related. This depends on the symmetry which connects the top and bottom surface in an electrostatically stable slab geometry. For $\mathrm{Cr_2O_3}$ the symmetry connecting top and bottom surfaces for any surface orientation is the product of inversion and time-reversal symmetry, the bulk $\mathcal{I}\mathcal{T}$ symmetry. This symmetry forces the opposite surfaces of $\mathrm{Cr_2O_3}$ slab to have opposite signs of surface magnetization and hence opposite exchange splitting (Fig.~\ref{fig:Cr2O3topbot}). This is shown explicitly for the $(001)$ surface in Fig.~\ref{fig:Cr2O3topbot}, but we emphasize that it holds for the $(100)$ and $(\bar{1}20)$ surfaces as well. As a result, the \textit{net} band structure of a $(001)$-oriented slab of $\mathrm{Cr_2O_3}$ is spin-degenerate and has zero net magnetization, as discussed previously by Tao et al.~\cite{taoLayerHallDetection2024}.

Nevertheless, the spin-splitting of a single surface in such a case can still be probed with surface-sensitive probes such as spin-polarized angle-resolved photoemission spectroscopy (sp-ARPES), or the magneto-optical Kerr effect (MOKE) for example. Spin-polarized current in such a case can also generated provided that opposite surfaces with opposite polarizations remain electronically decoupled, for examples if the surfaces of $\mathrm{Cr_2O_3}$ are doped into a metallic states while the bulk remains insulating.

\begin{figure}
    \centering
    \includegraphics[width=3.4in]{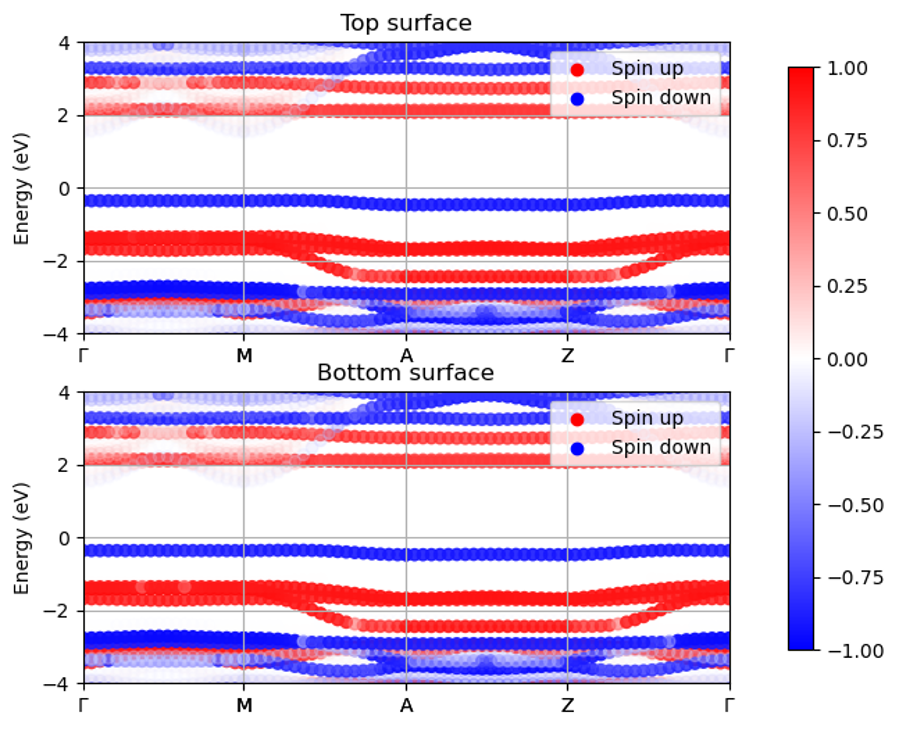}
    \caption{Spin projected at the top/bottom surface for a (110) slab geometry of $\mathrm{FeF_2}$. The values for the top and bottom surface are equal. Each surfaces referred to consist of 2 Fe sites and 3 F sites each.}
    \label{fig:FeF2topbot}
\end{figure}

On the other hand, for $(110)$ $\mathrm{FeF_2}$ the symmetry connecting the two surfaces is just the bulk symmetry inversion $\mathcal{I}$. As there is no time-reversal involved, the magnetization on both top and bottom surface has the same sign, meaning a slab of $\mathrm{FeF_2}$ has a small \textit{net} magnetization, and a corresponding \textit{net} exchange-splitting due to top and bottom surfaces having the same sign of spin-splitting(Fig.~\ref{fig:FeF2topbot}). Thus, for bulk AFMs lacking $\mathcal{I}\mathcal{T}$ symmetry but with pure inversion symmetry, we expect that spin-splitting and associated effects will be detectable and exploitable even via bulk techniques, due to the same-sign spin-splitting contributions of top and bottom surfaces.

\subsection{Magnetization canting at $\mathrm{Cr_2O_3}$ (100)} \label{AppendixCr2O3100canting}

\begin{figure}
    \centering
    \includegraphics[width=3.4in]{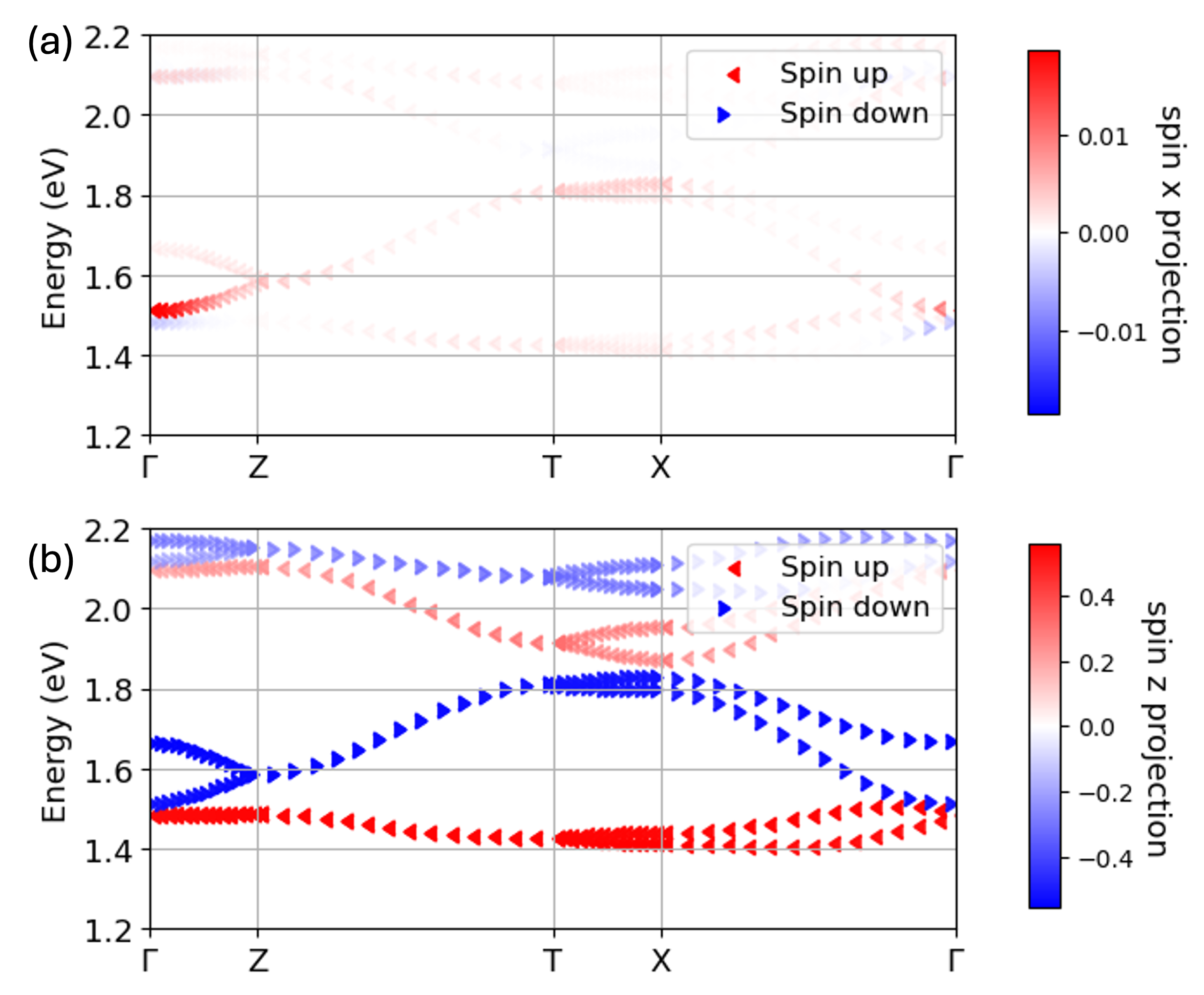}
    \caption{Spin projection with the spin-axis taken (a) out-of-plane (b) along the Néel vector. The canting of the magnetization at the surface was taken as $0.5^{\circ}$ degrees with respect to the Néel vector.}
    \label{fig:Cr2O3100canting}
\end{figure}

We justify our claim that the relativistic effects at the surface of $\mathrm{Cr_2O_3}$ are negligible for the overall band structure shown in Fig.~\ref{fig:Cr2O3bands}(c). The DFT calculations use the same settings as the ($\bar{1}20$) surface of $\mathrm{Cr_2O_3}$ except for the canting angle which is $0.5^{\circ}$ with respect to the bulk Néel vector in the out-of-plane direction. This canting is close to the optimal canting angle~\cite{weberSurfaceMagnetizationAntiferromagnets2024}. 
Fig.~\ref{fig:Cr2O3100canting} show the spin projection with the spin-axis taken (a) out-of-plane (b) along the Néel vector. The energy shift due to the additional spin-orbit coupling and magnetization is $\sim0.002\ \mathrm{eV}$, which is far smaller than the energy scale we point out in Fig.~\ref{fig:Cr2O3bands}(c). As an additional check we also consider the projection of the spin along the out-of-plane direction, similar to the ($\bar{1}20$) surface. In the energy window we consider the projection along the out-of-plane direction is significantly smaller than along the Néel vector also showing that the addition of SOC has a negligible effect on the band structure.

\FloatBarrier            
\nocite{*}
\bibliography{MyLibrary}
\end{document}